\documentstyle[epsfig,aps,prc]{revtex}
\begin{document}

\title{Universal trend of the information entropy of a fermion in a
mean field}

\author{C.P. Panos, S.E. Massen and C.G. Koutroulos}
\address{Department of Theoretical Physics,
Aristotle University of Thessaloniki,
54006 Thessaloniki, Greece }

\maketitle

\begin{abstract}

We calculate the information entropy of single-particle states in
position-space $S_{r}$ and momentum-space $S_{k}$ for a nucleon
in a nucleus, a $\Lambda$ particle in a hypernucleus and an electron
in an atomic cluster. It is seen that $S_{r}$ and $S_{k}$ obey the
same approximate functional form as functions of the number of
particles, $S_{r}$ ({\rm or} $S_{k}) = a+bN^{1/3}$ in all of the above
many-body systems in position- and momentum- space separately.
The net information content $S_{r}+S_{k}$ is a slowly varying
function of $N$ of the same form as above. The entropy sum $S_{r}+S_{k}$
is invariant to uniform scaling of coordinates and a characteristic
of the single-particle states of a specific system. The order of
single-particle states according to $S_r +S_k$ is the same as their
classification according to energy keeping the quantum number $n$ constant.
The spin-orbit splitting is reproduced correctly.
It is also seen that $S_{r}+S_{k}$ enhances with excitation of a fermion in
a quantum-mechanical system.
Finally, we establish a relationship of $S_r +S_k$ with the energy of the
corresponding single-particle state i.e. $S_r +S_k = k \ln (\mu E +\nu)$.
This relation holds for all the systems under consideration.
\end{abstract}

\section{Introduction}

The information entropy for a continuous probability distribution
$p(x)$ in one dimension is defined by the expression:
\begin{equation}
S=-{\int}p(x)\ln(p(x))dx
\label{1}
\end{equation}
where ${\int}p(x)dx=1$.

$S$ is measured in bits if the base of the logarithm is 2 and nats
(natural units of information) if the logarithm is natural. It
represents the information content of a probability distribution
as well as a measure of uncertainty of the corresponding state.
We note that the information and thermodynamic entropy are different
concepts but can be connected employing some assumptions.

An important step in the past was the discovery in [1]
of an entropic uncertainty relation (EUR), which for a three-dimensional
system has the form:
\begin{equation}
S_{r}+S_{k}{\ge} 3(1+\ln{\pi}) \cong 6.434,\quad (\hbar =1)
\label{2}
\end{equation}
where
\begin{equation}
S_{r}=-{\int}{\rho}(\vec{r}) \ln {\rho}(\vec{r})d{\vec{r}}
\label{3}
\end{equation}
is the information entropy in position-space
\begin{equation}
S_{k}=-{\int}n(\vec{k})\ln n(\vec{k})d{\vec{k}}
\label{4}
\end{equation}
is the information entropy in momentum-space and $\rho (\vec{r})$,
$n(\vec{k})$ are the density distributions in position- and momentum-
space respectively, normalized to unity.

The lower bound in (\ref{2}) is attained for gaussian density distributions.
The physical meaning of the above inequality is the following: an
increase of $S_{k}$ corresponds to a decrease of $S_{r}$ and vice
versa, which indicates that a diffuse density distribution $n(k)$
in momentum space is associated with a localised density distribution
$\rho (r)$ in configuration space and vice versa.

Relation (2) represents a strengthened version of Heisenberg's uncertainty
principle for two reasons: first EUR leads to Heisenberg's uncertainty
relation but the inverse is not true. Second the right-hand-side of
EUR does not depend on the state of the system, while in Heisenberg's
relation does depend.
It is obvious from (\ref{3}) and (\ref{4}) that $S_r$, $S_k$ depend on the
unit of length in measuring $\rho(\vec{r})$ and $n(\vec{k})$. However, the
important quantity is the entropy sum $S_r + S_k$ (net information content
of the state) which is invariant to uniform scaling of coordinates.

Information entropy was employed in the past for
the study of quantum mechanical systems [1-9].
Recently \cite{10} we studied the position- and momentum- space 
information entropies $S_{r}$, $S_{k}$ respectively for various 
systems: the nuclear density distribution of nuclei, the electron 
density distribution of atoms and  the valence electron density 
distribution of atomic clusters. We showed that a similar functional 
form $S=a+b \ln N$ for the entropy as function of the number of
particles $N$ holds approximately for the above systems. We
conjectured that this is a universal property of a many-fermion
system in a mean field.

The concept of information entropy proved also to be fruitful in a
different context \cite{9}. We used the formalism of Ghosh, Berkowitz and
Parr \cite{11} within the ground state density functional framework, to
define the concept of an information entropy associated with the
density distribution of a nuclear system. It turned out that $S$ increases
with the quality of the wave function and can serve as a criterion of
the quality of a nuclear model.

Another interesting result \cite{12} is the fact that the entropy of an
$N$-photon state subjected to Gaussian noise increases linearly with
the logarithm of $N$.

Encouraged by previous work we attempt in the present paper to calculate
$S_{r}$, $S_{k}$ for the wave functions of single-particle states (instead
of the total densities as in \cite{10}) for various systems i.e. a nucleon
in a nucleus, an electron in an atomic cluster and a $\Lambda$ particle
in a hypernucleus. We employ for these systems models existing in the
literature.

Our aim is to investigate the dependence of $S_{r}$, $S_{k}$ on the
excitation of a fermion in a quantum-mechanical system as well as its
dependence on the system under consideration and the number of the
particles $N$.
We also attempt to connect the information entropy with the energy
of the single-particle state.
The study of the dependence of $S$ on the quantum state of a system
is also interesting (as stated in \cite{7}) for two reasons: (i) The
information-theoretical and physical entropy are connected via
Boltzmann's constant $k_B$ by the Jayne's relation $S_{phys}=k_B S_{inf}$.
Thus one can ascribe to any quantum object a certain value of its
physical entropy $S_{phys}$ if one calculates $S_{inf}$.
(ii) It is interesting to know the value of the information entropy
which is a measure of the spatial "spreading out" of the wave function for
various states of various systems.

The present paper is organized as follows: In sec. 2 we calculate $S_{r}$, 
$S_{k}$ for single-particle states of a nucleon in a nucleus as function of 
the number of nucleons $N$ using the simplest model available i.e. the
harmonic oscillator potential and a more realistic one (Skyrme). In sec. 3
we calculate $S_{r}$, $S_{k}$ for a $\Lambda$ particle in a hypernucleus 
employing a simple and (semi-) analytical relativistic model. In sec 4 we 
determine $S_{r}$, $S_{k}$ for the single-particle states of an electron 
in atomic (metallic) clusters using the Woods-Saxon potential. In sec. 5 
we present a relationship of $S_r + S_k$ with the energy. Finally, sec. 6
contains a discussion of our results (comparison of sec. 2, 3 and 4) and
our conclusions.

\section{Information entropy for a nucleon in a nucleus}

The information entropy $S_{r}$ in position-space for a single-particle
wave function $\psi (\vec{r})$ is defined as
\begin{equation}
S_{r}=-{\int}|\psi (\vec{r})|^{2} \ln |\psi (\vec{r})|^{2} d{\vec{r}}
\label{5}
\end{equation}
while the entropy $S_{k}$ in momentum-space is
\begin{equation}
S_{k}=-{\int}|\phi (\vec{k})|^{2} \ln |\phi (\vec{k})|^{2}d{\vec{k}}
\label{6}
\end{equation}
where $\phi (\vec{k})$ is the Fourier transform of $\psi (\vec{r})$.

In this section we calculate $S_{r}$ and $S_{k}$ for the single-particle
states $1s$, $1p$, $1d,\cdots$ of a nucleon in a nucleus in the framework 
of the harmonic oscillator (HO) model. We use for the HO parameter the 
well-known expression $\hbar \omega=41A^{-1/3}\ $MeV.

We find that the value of ${\hbar}{\omega}$ is important only for
$S_{r}$, $S_{k}$, while the net information content $S=S_{r}+S_{k}$
is independent of $\hbar \omega$ and consequently of $A$. It depends only 
on the state under consideration and characterises it. 
These values for the states $1s$, $1p$, $1d$ and $2s$ are
6.4341, 7.8388, 8.6651 and 8.3015 respectively.

However, the HO model is a simplification. Thus, we employed a more 
realistic parametrization of the nuclear mean field i.e. the Skyrme 
(Sk III) interaction \cite{13}.
In this model protons and neutrons move in different potentials. We choose
to work with protons. However, similar results can be obtained for neutrons.
We found that the values for $S_{r}$, $S_{k}$ obtained from the wave 
functions of single-particle states calculated according to Sk III are 
represented well by the expression
\begin{equation}
S_{r}\, ({\rm or} \, S_{k}) =a+bN^{1/3} 
\label{7}
\end{equation}
while  $S_r +S_k$ is a slowly varying function of $N$ of the same form as
(\ref{7}).  The values of the parameters are shown in Table 1.

In Fig. 1a we plot our fitted expressions (Sk III) $S_r$ 
$({\rm or}\, S_k) =a+bN^{1/3}$ for the entropies $S_{r}$, $S_{k}$, 
$S_{r}+S_{k}$ of $1s$-states as functions of $N^{1/3}$. The lines 
correspond to our fitted expressions, while the corresponding values of 
our numerical calculations are denoted by squares for $S_r$, circles 
for $S_k$ and triangles for $S_r+S_k$.
Similar graphs can be plotted for the higher states $1p, 1d, 2s,\cdots$. 
From Fig. 1a we see that the values of the entropies are represented well 
by our fitted expressions. In Fig. 1b we compare the sum 
$S_{r}+S_{k}=a+bN^{1/3}$ for various single-particle states. We observe 
that the entropy sum $S_{r}+S_{k}$ enhances with the excitation of the 
single-particle states. We see that $S_{r}+S_{k}$ is a slowly varying 
function of $N$. We also note that the spin-orbit splitting is reproduced 
correctly i.e. the state $1p_{3/2}$ is lower than $1p_{1/2}$ e.t.c. 
(as for the energy) although their difference is small and 
cannot be shown in the figure.

\section{Information entropy for a $\Lambda$ in a hypernucleus}

We employ a simple and (semi-) analytical relativistic model of a
hypernucleus of ref. \cite{14,15}, where the Dirac equation with a scalar
potential $U_{S}(r)$ and the fourth component of a vector potential
$U_{V}(r)$ was considered in the case of rectangular shapes of these
potentials with the same radius:
$$
R=r_{0}A_{core}^{1/3}
$$
In \cite{14} the Dirac equation was solved and gave the wave functions $G(r)$
and $F(r)$ for the large and small components for a $\Lambda$ particle
in a hypernucleus. These components can be found in relations (\ref{10}) and 
(\ref{11}) of \cite{14}.

The Dirac spinors in terms of large $(G)$ and small $(F)$ components
can be expressed:
\begin{equation}  
\psi=\left(\begin{array}{c}
                iG(r)/r \cr
                F(r)/r 
              \end{array}\right)
\label{8}
\end{equation}

The density distribution of a $\Lambda$ in position space is:
\begin{equation}
\rho(r)=\frac{1}{4\pi}[G^{2}(r)/r^{2} + F^{2}(r)/r^{2} ]
\label{9}
\end{equation}
and the normalization is 
$$
4{\pi} {\int}^{\infty}_{0}{\rho}(r)r^{2}dr=1
$$
In momentum-space we have:
\begin{equation}
{\phi}(k)=\left(\begin{array}{c}
                      iX(k) \cr
                      Y(k) 
                     \end{array}\right)
\label{10}
\end{equation}
where $X(k)$ and $Y(k)$ are the Fourier transforms of $G(r)/r$ and
$F(r)/r$ respectively. Thus the density distribution in momentum-space
is given by:
\begin{equation}
n(k)=\frac{1}{4\pi}[X^{2}(k)+Y^{2}(k)]
\label{11}
\end{equation}
and the normalization is:
$$
4{\pi}{\int}^{\infty}_{0}n(k)k^{2}dk=1
$$
The information entropies of the $\Lambda$ particle are calculated
according to the relations:
\begin{equation}
S_{r}=-4 \pi \int \rho (r) \ln \rho (r) r^{2} dr
\label{12}
\end{equation}
\begin{equation}
S_{k}=-4{\pi}{\int}n(k) \ln n(k)k^{2}dk 
\label{13}
\end{equation}
where $\rho (r)$ and $n(k)$ are given by (\ref{9}) and (\ref{11}) 
respectively.

For the depths of the potential we used the values \cite{14}:
$D_{+}=30.55$ MeV, $D_{-}=300$ MeV, $r_{0}=1.01$ fm
and the radius parameter $R=r_{0}A_{core}^{1/3}$ obtained by
fitting the experimental binding energies of the ground state
of the $\Lambda$ particle. In the following we put 
$A_{core}=N=$ number of particles.

Next we fitted the expressions $S_{r}$ $({\rm or}\, S_{k}) =a+bN^{1/3}$ to 
the values of $S_{r}$, $S_{k}$ calculated from (\ref{12}) and (\ref{13}) 
and found that these values are represented well. The values of the 
parameters $a$ and $b$ for various states are shown in Table 1.

In Fig. 2a we plot our fitted expressions for $S_{r},S_{k},S_{r}+S_{k}$ 
as functions of $N^{1/3}$ for the $1s$ state. This is done for a $\Lambda$ 
in a hypernucleus in a similar way as for a nucleon in a nucleus (Fig 1a). 
Similar graphs can be plotted for the higher states. In fig. 2b we 
compare the sum $S_{r}+S_{k}$ for various single-particle states
of a $\Lambda$ (similar with Fig 1b). The spin orbit splitting is
reproduced correctly as in nuclei (Sec. 2).

\section{Information entropy for an electron in an atomic cluster.}

We consider atomic (metallic) clusters composed of neutral sodium
atoms, where the electrons move in an effective radial electronic
potential parametrized by Woods-Saxon potential of the form:
\begin{equation}
V_{WS}(r)=\frac{-V_{0}}{1+\exp [(r-R)/a]}
\label{WS}
\end{equation}
with $V_{0}=6\ $ eV, $R=r_0 N^{1/3}$, $r_{0}=2.25\ \AA$ and
$a=0.74\ \AA$. For a detailed study regarding the parametrization
of Ekardt's potentials see ref \cite{16}.

We found the wave functions of the single-particle states in configuration
space and by  Fourier transform the corresponding ones in momentum
space by solving numerically the Schr{\"{o}}ndinger equation for atomic
clusters for various values of the number of valence electrons $N$.
Using the above wave functions, we calculated the information entropies
$S_{r}$, $S_{k}$ (relations (\ref{5}) and (\ref{6}) ) for the 
single-particle states instead of the total density distributions as in 
ref. \cite{10}. Then we fitted the form $S_r$ $({\rm or}\, S_k)=a+bN^{1/3}$ 
to these values and found that these expressions represent 
well the values of $S_{r}$,$S_{k}$. In Fig. 3a we plot $S_{r}$, $S_{k}$ and 
$S_{r}+S_{k}$ as functions of $N^{1/3}$ (similar as Fig. 1a, 2a) and in
Fig. 3b we compare $S_{r}+S_{k}$ for various states (similar as in Fig.
1b, 2b). In Table 1 we present the values of the parameters $a$ and $b$
which were obtained from the fitting.

\section{Relationship of the information entropy with the energy of
single-particle states}

In Fig. 4 we plot $S_r +S_k$, obtained with the HO model of the nucleus,
versus the energy of the single-particle states. We use 
$\hbar \omega = 41 A^{-1/3}$ with $A=208$ (Pb) and keep the quantum number 
$n$ equal to 1. A fitting procedure gives for $n=1$, the relation:
\begin{equation}
S=k \ln (\mu E + \nu )
\label{S(E1)}
\end{equation}
where $k=2.0206$, $\mu =3.5373\ {\rm MeV}^{-1}$ and $\nu=-12.5320$. Similar 
relations hold for $n > 1$.

Next we plot the sum $S_r + S_k$ as function of the energy $E$ of 
single-particle states for a proton in a nucleus according to Sk III 
interaction for ${}^{208}$Pb (Fig. 5) and an electron in atomic cluster 
with $N=198$  (Fig. 6) for $n=1$.
Similar curves hold for higher values of $n > 1$. In both cases
the dependence of $S_r + S_k$ on $E$ can be represented well by the functional
form (\ref{S(E1)}). The values of the constants are the following
\[
k=1.5262, \quad \mu=17.3043\ {\rm MeV}^{-1}, \quad 
\nu=793.109 \quad {\rm for\, a\, proton\, in\, a\, nucleus}
\]
\[
k=1.2386, \quad \mu=1481.48\ {\rm eV}^{-1}, \quad 
\nu=8730.52 \quad {\rm for\, an\, electron\, in\, a\,
cluster}
\]

 A similar relation may be obtained for a $\Lambda$ in a
hypernucleus but the number of values of $S_r + S_k$ available is small.
It is the first time in our research on information entropy that we observe
such a relationship of $S_r + S_k$ with a fundamental quantity as the energy.
We note that in Fig. 5 there are pairs of points with almost the same
$S_r + S_k$ which reproduce the spin-orbit splitting.

\section{Discussion and Conclusions}

Comparing our results in sections 2, 3, and 4, we see that a similar
functional form $S_r$ $({\rm or}\, S_k) =a+bN^{1/3}$ describes well the 
information entropies $S_{r}$, $S_{k}$ of the single-particle states for a
nucleon in a nucleus, a $\Lambda$ in a hypernucleus and a valence electron 
in an atomic cluster, although the single-particle potentials are different. 
We conjecture that this is a universal trend of the information entropies
$S_{r}$, $S_{k}$ for a fermion in a mean field, while the net information
content $S_{r}+S_{k}$ of the single-particle states of a fermion in
a mean field is a slowly varying function of $N$ of the form $S=a+bN^{1/3}$
for the systems considered above. For nuclei and the simple HO potential
$S_{r}+S_{k}$ is exactly a constant independent of $N$ i.e. $b=0$. 

We note that in \cite{10} we found the universal property $S=a+b \ln N$ 
for the total density distributions of various systems.

In both cases it is not clear why $S$ depends 
linearly on $\ln N$
(total densities) or linearly on $N^{1/3}$ (single-particle states)
but we note that in atomic physics there is already a connection of
the information entropy with experiment i.e. with fundamental and/or
experimental quantities e.g. the kinetic energy or the magnetic
susceptibility. Both characteristics have been used in the study
of the dynamics of atomic and molecular systems \cite{17}. This connection
established the information entropy as an interesting entity for atomic
physics. 
In the present paper we obtained a relationship of $S_r + S_k$ with a
fundamental quantity as the energy of the single-particle states, i.e.
$S=k \ln (\mu E + \nu )$. It is remarkable that the same functional form 
holds for various systems.

\newpage
\begin{figure}
\label{fig-1}
\begin{center}
\begin{tabular}{cc}
{\epsfig{figure=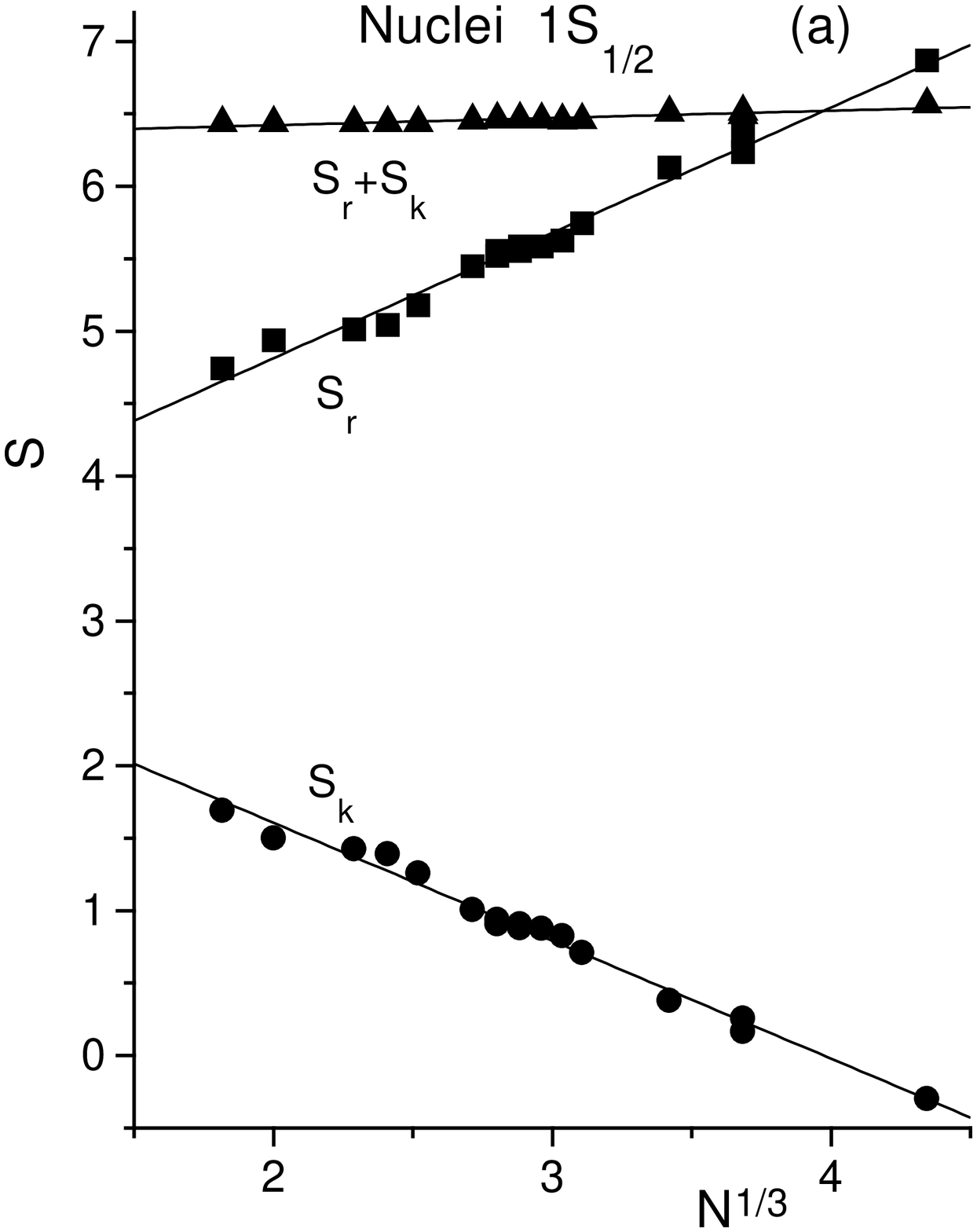,width=6.cm} } &
{\epsfig{figure=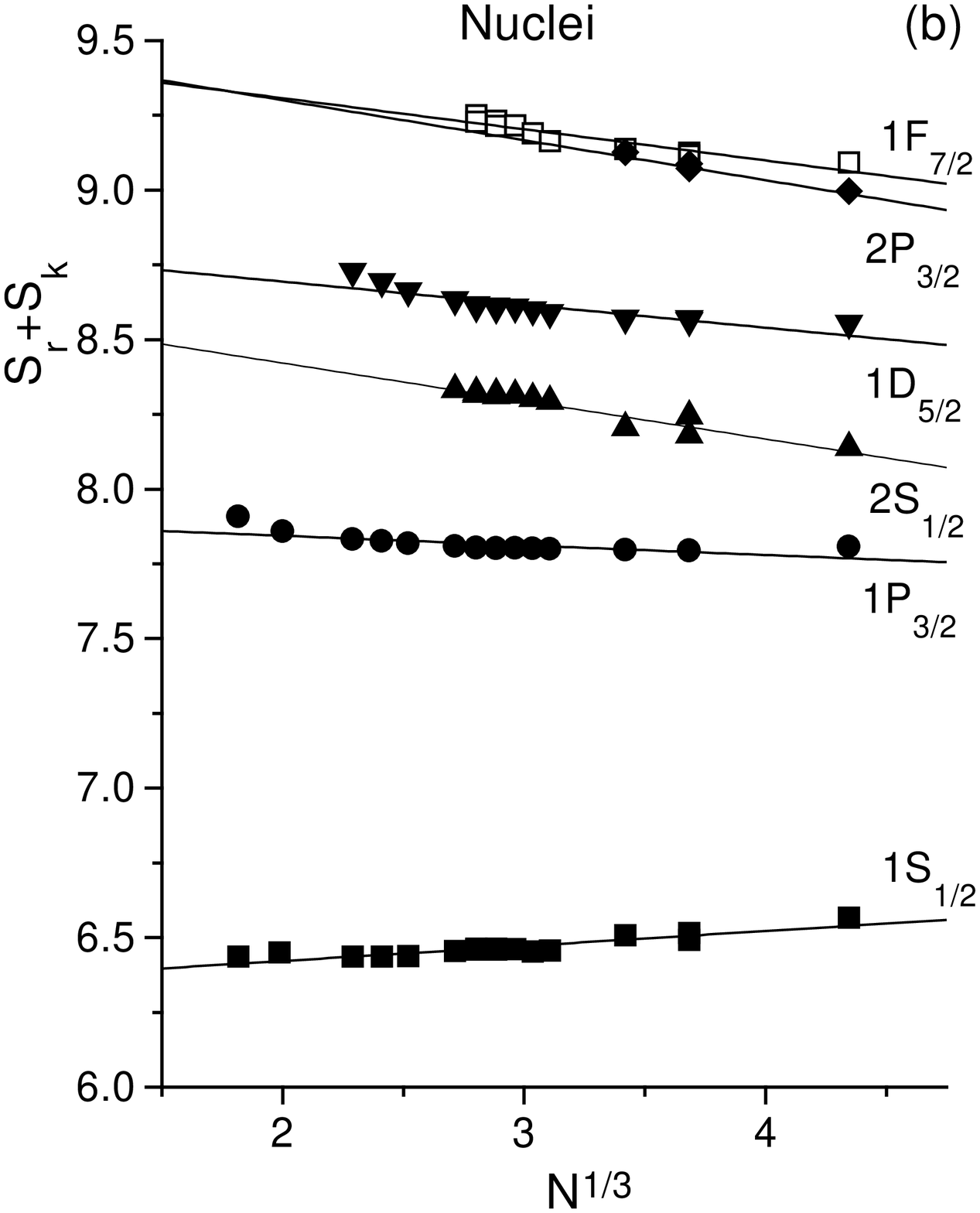,width=6.cm} } 
\end{tabular}
\end{center}
\caption{ {\bf (a)} Values of the information entropies $S_r$ (squares),
$S_k$ (circles) and $S_r+S_k$ (triangles), calculated numerically, versus
the number of particles $N$. These values correspond to the single-particle
states of a proton in various nuclei, according to the Sk III interaction.
The lines correspond to our fitted expressions $S_r$ 
$({\rm or}\, S_k) =a+b N^{1/3}$.
{\bf (b)} Comparison of the sum $S_r+S_k$ for various proton single-particle
states.}
\end{figure}

\begin{figure}
\label{fig-2}
\begin{center}
\begin{tabular}{cc}
{\epsfig{figure=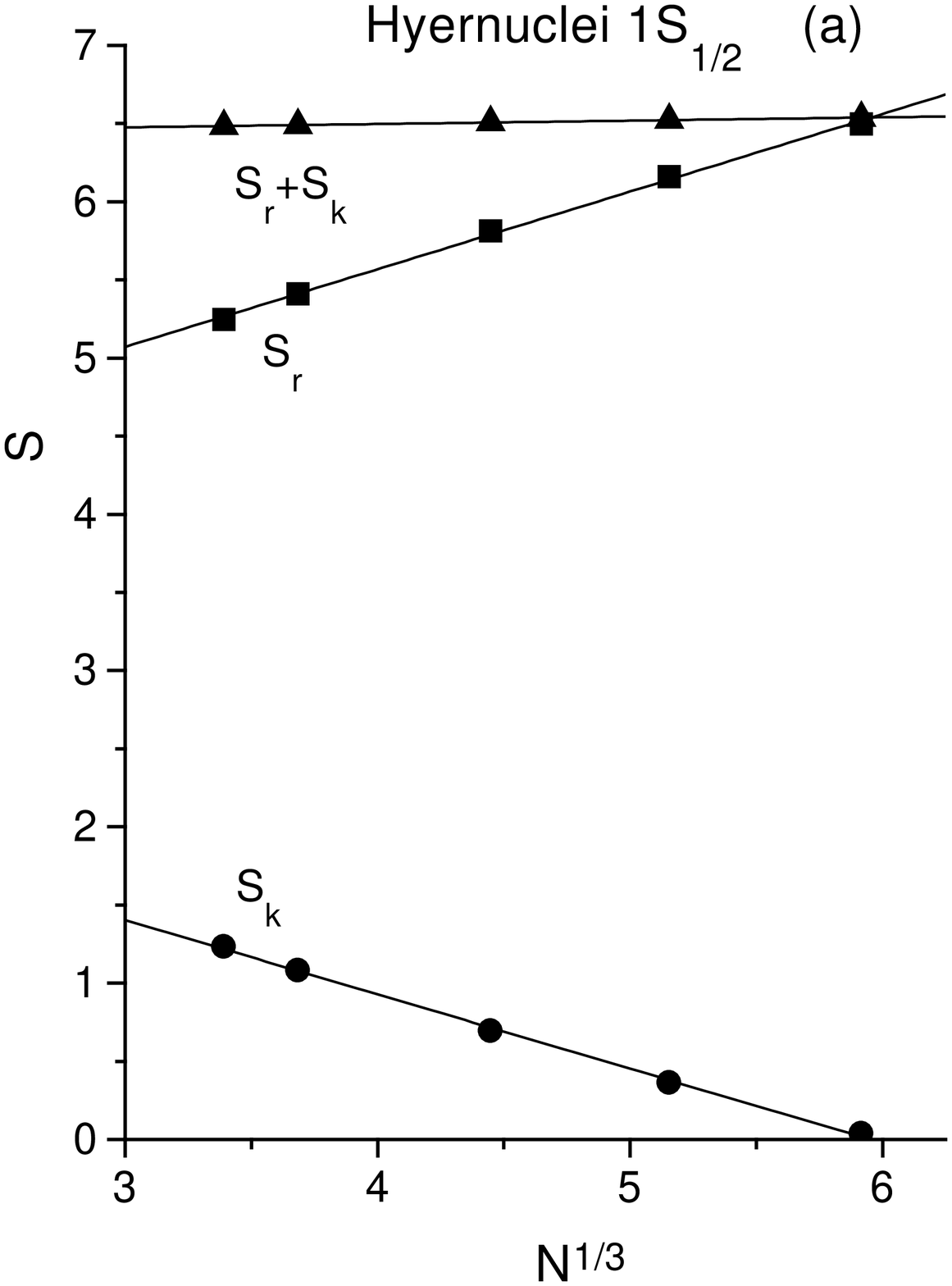,width=6.cm} } &
{\epsfig{figure=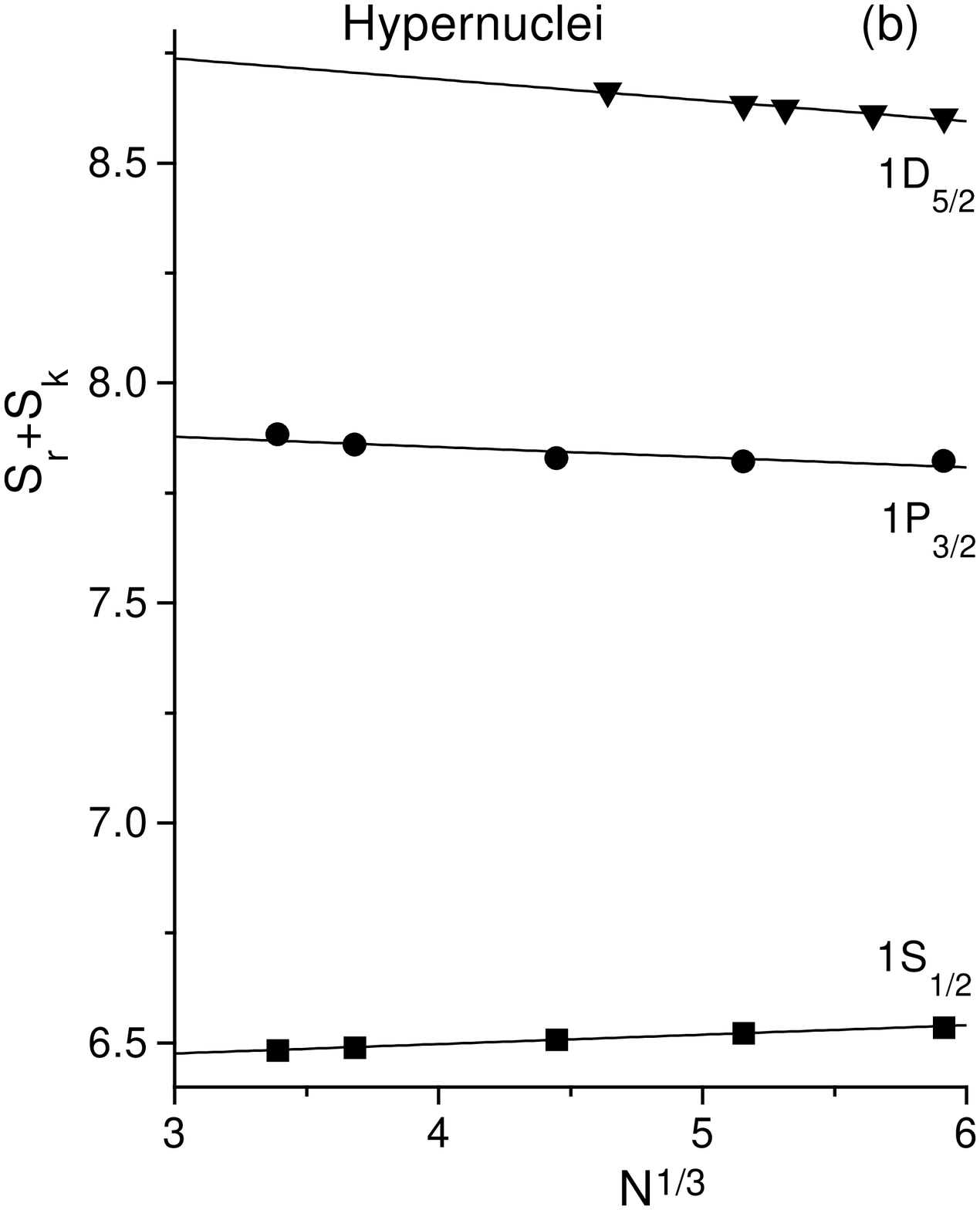,width=6.cm} }
\end{tabular}
\end{center}
\caption{The same as in Fig. 1 for a $\Lambda$ in hypernuclei 
employing a relativistic model.}
\end{figure}

\begin{figure}
\label{fig-3}
\begin{center}
\begin{tabular}{cc}
{\epsfig{figure=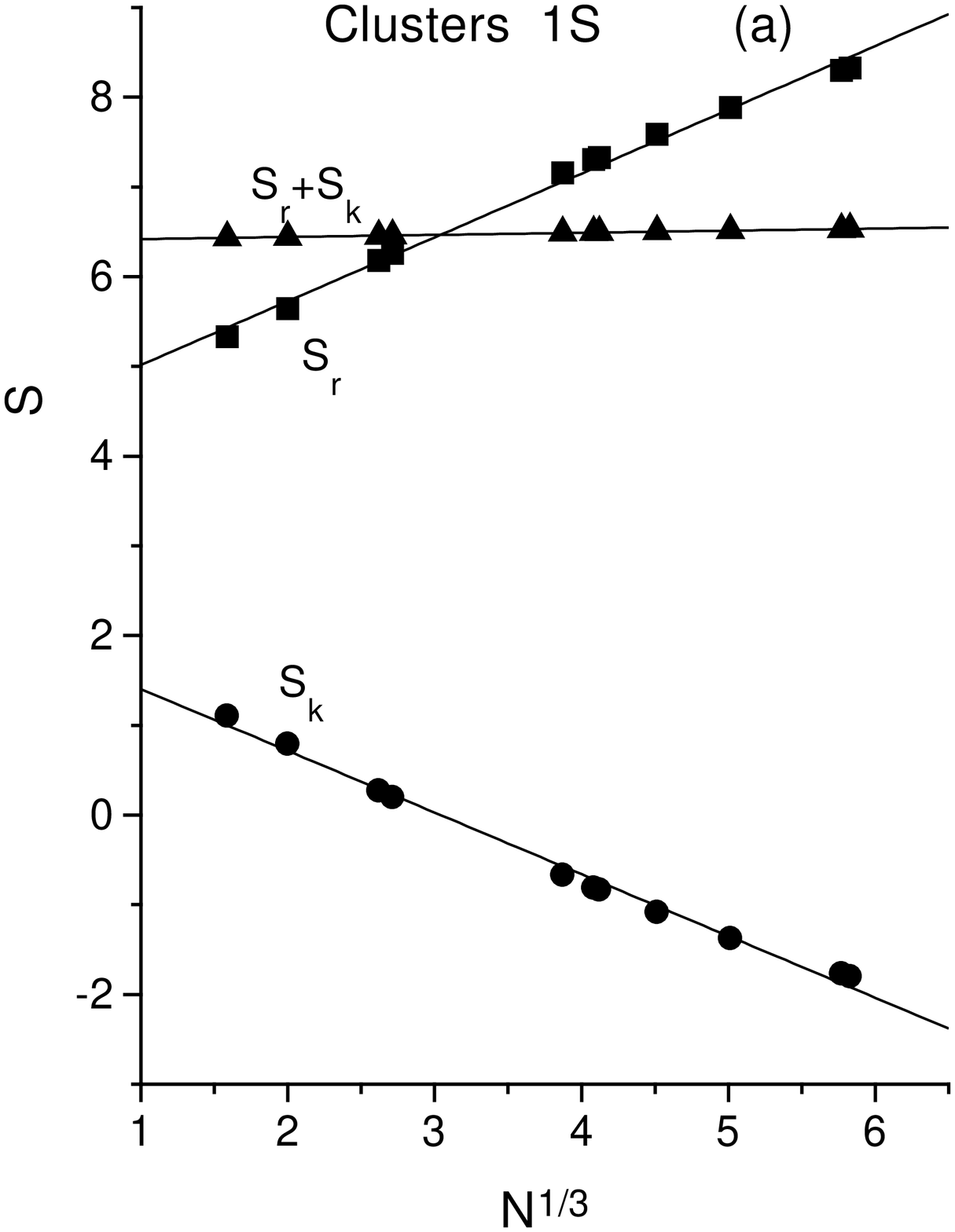,width=6.cm} } &
{\epsfig{figure=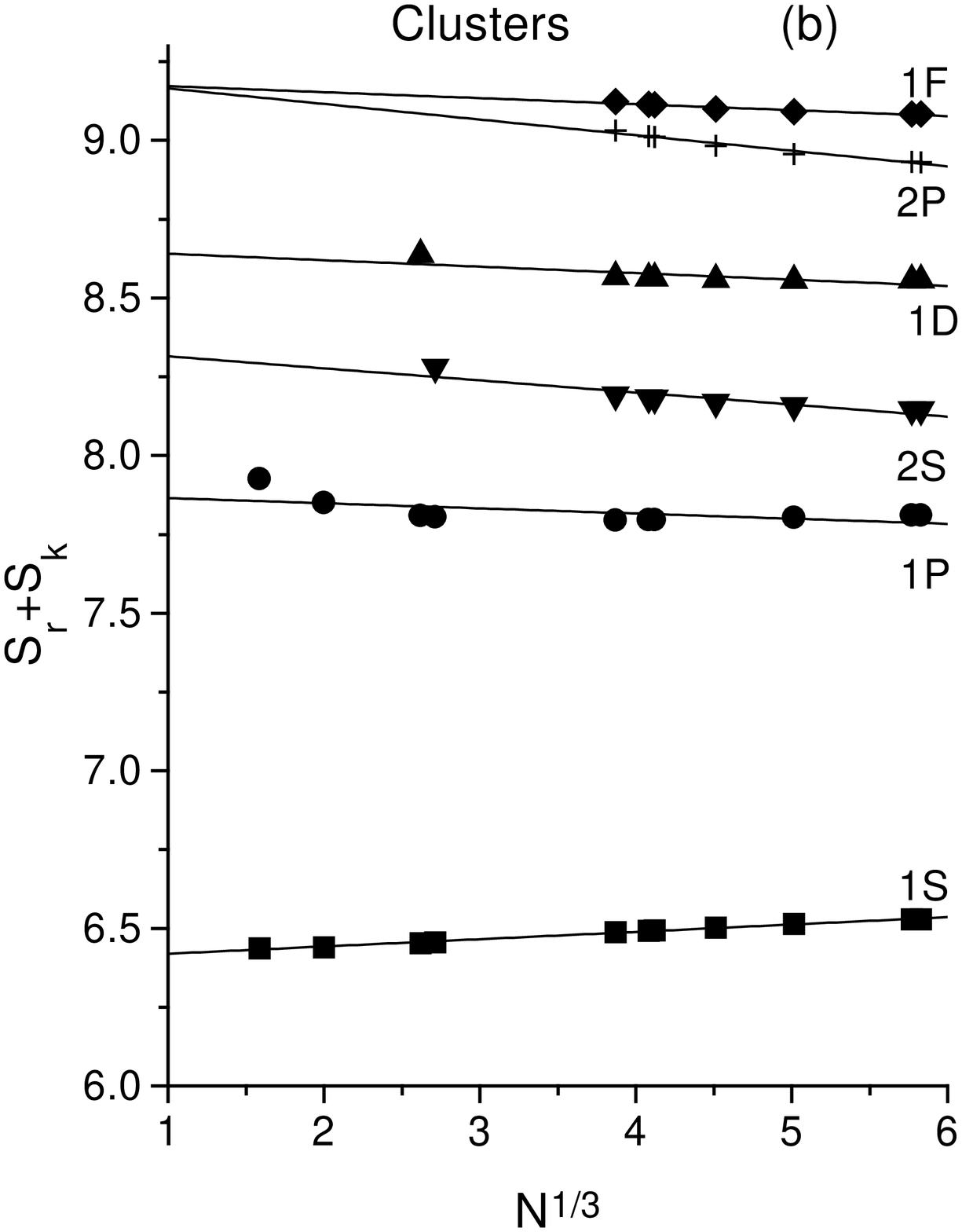,width=6.cm} } 
\end{tabular}
\end{center}
\caption{The same as in Fig. 1 for an electron in atomic clusters with 
the Woods-Saxon potential.}
\end{figure}

\begin{figure}
\label{fig-4}
\centerline{\epsfig{figure=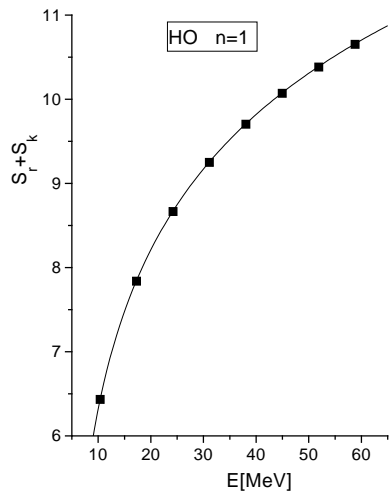,width=6.cm} }
\caption{The values of the energy $E$ (squares) of single-particle
states for a nucleon in Pb$^{208}$ according to the HO model for $n=1$. The
line corresponds to our fitted expression $S=k \ln (\mu E +\nu)$.}
\end{figure}

\begin{figure}
\label{fig-5}
\centerline{\epsfig{figure=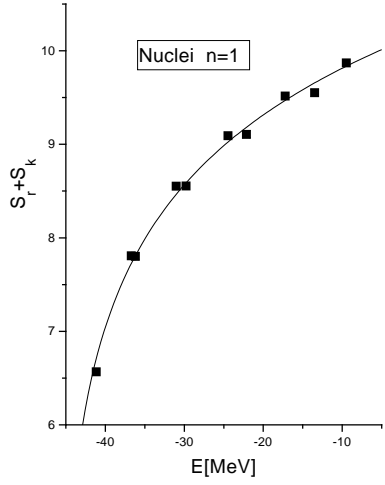,width=6.cm} }
\caption{The same as in Fig. 4 but for a proton in Pb$^{208}$ according to 
the Sk III interaction for $n=1$.}
\end{figure}

\begin{figure}
\label{fig-6}
\centerline{\epsfig{figure=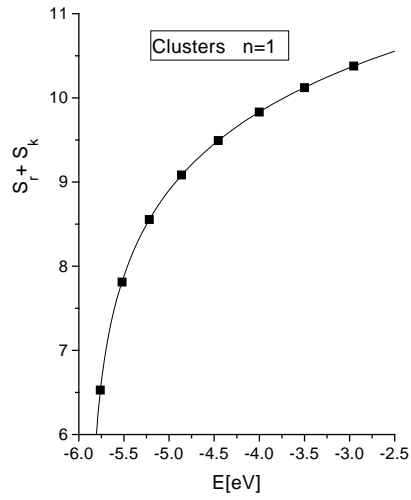,width=6.cm} }
\caption{The same as in Fig. 4 but for an electron in atomic clusters (N=198)
using the Woods-Saxon potential for $n=1$.}
\end{figure}

\newpage
\begin{table}
\caption{ 
Values of the parameters $a$ and $b$ which appear in the expressions $S_r$ 
$({\rm or}\, S_k) =a+b N^{1/3}$ for a nucleon (proton) in nuclei according to 
the Sk III interaction, a $\Lambda$ in hypernuclei according to a relativistic 
model and  an electron in atomic clusters with Woods-Saxon potential.}

\begin{center}
\begin{tabular}{l  c c c c c c c }
\hline
Case&State    & \multicolumn{2}{c}{$S_r$}& \multicolumn{2}{c}{$S_k$} & 
      \multicolumn{2}{c}{$S_r+S_k$}\\
 &   & $a$ & $b$ & $a$ & $b$ & $a$ & $b$  \\
\hline
 &   &   &   &   &   &   & \\
Nucleus&$1s_{1/2}$      &3.0831& 0.8652&3.2353&-0.8140&6.3217& 0.0501\\ 
       &$1p_{3/2}$& 4.2824& 0.6368& 3.6256&-0.6688& 7.9084&-0.0322\\  
       &$1p_{1/2}$& 4.2724& 0.6235& 3.6799&-0.6675& 7.9521&-0.0439\\
       &$1d_{5/2}$& 4.7500& 0.5743& 4.0980&-0.6513& 8.8480&-0.0771\\
       &$1d_{3/2}$& 4.9553& 0.5042& 4.0080&-0.6071& 8.9618&-0.1024\\  
       &$2s_{1/2}$& 5.2456& 0.3364& 3.4355&-0.4641& 8.6756&-0.1260 \\
 &   &   &   &   &   &   & \\
Hyper- &$1s_{1/2}$& 3.5817& 0.4967& 2.8303&-0.4756& 6.4120& 0.0214\\ 
nucleus&$1p_{3/2}$& 4.4347& 0.3789& 3.5123&-0.4021& 7.9475&-0.0232\\  
       &$1p_{1/2}$& 4.3764& 0.3835& 3.6575&-0.4199& 8.0342&-0.0364\\
       &$1d_{5/2}$& 5.2553& 0.2462& 3.6249&-0.2938& 8.8803&-0.0475\\
       &$1d_{3/2}$& 4.9910& 0.2819& 4.0503&-0.3543& 9.0414&-0.0724\\  
  &  &   &   &   &   &   &   \\
Cluster&$1s$& 4.3038& 0.7113 &2.0923& -0.6883& 6.3960&  0.0232 \\ 
       &$1p$& 5.1114& 0.6135 &2.7700& -0.6299& 7.8816& -0.0163 \\  
       &$1d$& 5.5191& 0.5636 &3.1420& -0.5842& 8.6611& -0.0205 \\
       &$2s$& 5.3858& 0.4918 &2.9672& -0.5301& 8.3536& -0.0383 \\
  &  &   &   &   &   &   &   \\
\hline
\end{tabular}
\end{center}
\end{table}

\begin{thebibliography}{99}
\bibitem{1} I. Bialynicki-Birula, J.Mycielski, Commun. Math. Phys. {\bf{44}}
 (1975) 129.

\bibitem{2} S.R. Gadre, Phys. Rev. {\bf{A 30}} (1984) 620.

\bibitem{3} S.R. Gadre, S.B. Sears, S.J. Chacravorty, R.D. Bendale,
 Phys. Rev. {\bf{A 32}} (1985) 2602.

\bibitem{4} S.R. Garde, R.D. Bendale, Phys. Rev. {\bf{A 36}} (1987) 1932.

\bibitem{5} M. Ohya, P. Petz, "Quantum entropy and its use" (Springer Berlin, 
1993).

\bibitem{6} A. Nagy, R.G. Parr, Int. J. Quant. Chem. {\bf{58}} (1996) 323.

\bibitem{7} V. Majernic, T. Opatrny, J. Phys. {\bf{A 29}} (1996) 2187.

\bibitem{8} C.P. Panos, S.E. Massen, Int.J. Mod. Phys. {\bf{E 6}} (1997) 497.

\bibitem{9} G.A. Lalazissis, S.E. Massen, C.P. Panos, S.S. Dimitrova, Int. J.
Mod. Phys. {\bf{E 7}} (1998) 485.

\bibitem{10} S.E. Massen, C.P. Panos , Phys. Lett. {\bf{A 246}} (1998) 530.

\bibitem{11} S.K. Ghosh, M. Berkowitz, R.G. Parr, Proc. Natl. Acad. Sci. USA
{\bf 81} (1984) 8028.

\bibitem{12}M. Hall, Phys. Rev. {\bf{A 50}} (1994) 3295.

\bibitem{13} C.B. Dover, N.Van Giai, Nucl. Phys. {\bf{A 190}} (1972) 373.

\bibitem{14} M.E. Grypeos, C.G. Koutroulos and G.J. Papadopoulos, Phys. Rev.
{\bf{A 50}} (1994) 29.

\bibitem{15} G.J. Papadopoulos, C.G. Koutroulos and M.E. Grypeos, Int. J. Theor.
Phys. {\bf{39}} (2000) 455.

\bibitem{16} B.A.Kotsos, M.E. Grypeos, Physica {\bf{B 229}} (1997) 173.

\bibitem{17} R.G. Parr, W.Yang, "Density functional theory of atoms and 
molecules
(Oxford Univ. Press, New York, 1989).
\end{thebibliography}
\end{document}